\documentclass[twocolumn,twoside]{revtex4}
\usepackage{graphicx}
\usepackage{fancyhdr}
\begin{document}

\title{Exotic quarkonium physics prospects at Belle~II}

\author{J. V. Bennett}
\affiliation{The University of Mississippi, Oxford, MS, USA, 38655}

\begin{abstract}
The Belle~II experiment at the SuperKEKB energy-asymmetric $e^{+}e^{-}$ collider is a substantial upgrade of the B~factory facility 
at KEK in Tsukuba, Japan. It aims to record a factor of 50 times more data than its predecessor. The experiment completed a 
commissioning run in 2018, and began full operation in early 2019. Belle II is uniquely capable of studying the so-called ``XYZ'' particles: 
heavy exotic hadrons consisting of more than three quarks. First discovered by Belle, these now number in the dozens, and represent the 
emergence of a new category within quantum chromodynamics. This talk will present the prospects of Belle~II to explore exotic quarkonium 
physics.
\end{abstract}

\maketitle

\thispagestyle{fancy}


\section{Introduction}

The recent era of particle discoveries includes highly anticipated particles like the Higgs boson, but in terms of quantity is dominated by 
completely unexpected states that appear to defy the traditional paradigm of hadronic states being either mesonic ($q\bar{q}$) or baryonic 
($qqq$) in nature. Intriguingly, these ``exotic'' hadrons, so called for breaking the aforementioned paradigm, appear to be broadly interrelated, 
yet there is as of yet no consensus on a universal theoretical principal that explains them all. Complicating matters is the fact that many of 
these states have only been observed in a single production mechanism and often in only a single decay channel. A systematic search for 
additional production mechanisms and decays of these exotic candidate states, often collectively labelled ``XYZ states,'' are vital to developing 
a unifying theory and to further probe their interrelatedness. Despite the abundance of recent discoveries in hadron spectroscopy, the 
experimental effort remains very strong.

More than 30 exotic candidate states have been observed since the discovery of the $X(3872)$ in 2003~\cite{C03}, with new discoveries being made 
nearly every year. However, the experimental picture is muddied by the lack of experimental confirmation. Many XYZ states have been observed 
in a single decay mode and often at a single experiment. In order to develop a model to describe the XYZ spectrum, analogous to the quark model 
for the many hadronic states discovered in the 50s and 60s, it will be important to characterize the experimentally observed states by determining 
their quantum numbers and otherwise attempting to infer which are truly exotic and what quark configuration provides the dominant contribution. It 
will also be important to search for new states, which almost certainly exist, and compare and contrast the XYZ states in the charm and beauty 
sectors and across different experiments and production and decay mechanisms.  The Belle~II experiment will leverage the superior detector 
performance and much larger data sample, relative to its B~factory predecessors, to make extensive studies of XYZ states. While several 
experiments have the potential to make contributions to this effort, the Belle~II experiment is uniquely positioned to make significant contributions. 
Prospects for studies of XYZ states at Belle~II are presented here.

\section{Models for quarkonium-like states}

Some hadronic states are easily recognizable as exotic as they have properties directly indicative of degrees of freedom beyond those 
accessible by conventional $q\bar{q}$ or $qqq$ states. Others are deemed exotic due to an overpopulation of discovered states relative to
those expected from theoretical predictions. Collectively, these states provide the first possibility to explore nonstandard hadron configurations
that have long been conjectured~\cite{GM64}, including hybrids and multi-quark states. Most models of quarkonium-like states can be 
classified according to quark clustering and degrees of freedom. Several candidate models are briefly reviewed here. Of course, states are often
described as some combination of quark configurations.
 
Proposed more than four decades ago, the molecular picture, perhaps the most obvious, describes a tetraquark state as a hadronic molecule of 
two mesons, each consisting of a tightly bound quark anti-quark pair, weakly bound within a strong interaction volume~\cite{V76, R77}. Supporting 
this model is the evidence that many exotic candidates lie very near two-meson thresholds, such as the $X(3872)$~\cite{C03}. The lack of evidence 
for a prominent state near the $D^{0}\bar{D}^{0}$ threshold supports this model since the binding in the molecular picture is accomplished by pion 
exchange and angular momentum and parity restrictions prevent a three pseudoscalar coupling like $D^{0}\bar{D}^{0}\pi$~\cite{P08} However, the 
production rate of $X(3872)$ in prompt production, in which the meson pairs would be unlikely to have low enough relative momentum to bind, 
contradicts the molecular picture~\cite{A04, C13}. In addition, some states, such as the $Z_{c}(3900)$, are unlikely to be bound states of mesons 
since they lie above the threshold of meson pairs to which they decay~\cite{A13, L13}.

The hadrocharmonium picture contrasts with the molecular picture by describing the states as a heavy quark anti-quark core, about which the light 
quark $q\bar{q}$ forms~\cite{D08}, and was motivated by the strong preference for some exotic candidates to decay to conventional charmonium, 
rather than open flavor hadrons. From heavy-quark spin symmetry, one would expect the characteristics of the heavy quark anti-quark bound state 
to transfer to the charmonium final state. However, it is not clear why, if the binding energy is strong enough to hold states together long enough to 
be considered distinct, the quarkonium and light components would not immediately rearrange into two heavy open-flavor hadrons.

Intriguingly, another picture that describes exotic candidates actually predicts more states than have been observed. The diquark picture, in which 
two quarks or anti-quarks bind rather than quark anti-quark pairs, also benefits from a potential resolution of the stability problem mentioned for 
hadroquarkonium above, in that the diquark-antidiquark pair may be created with a large relative momentum that can only hadronize through the 
long distance tails of meson wave functions~\cite{B14}. The apparent connection between the $Y(4260)$ and $X(3872)$~\cite{A14} supports the 
diquark picture, which describes the Y states as orbital excitations of lower states~\cite{M14}.

The fact that many candidate exotic states lie just above thresholds suggests that some of the experimentally observed enhancements may be due 
to threshold rescattering rather than quark-level dynamics. The threshold ``cusps''~\cite{W48} produce phase motion that is similar to that of a 
Breit-Wigner amplitude~\cite{B08}.

\section{The Belle~II experiment}

The first generation B~factory experiments were a resounding success, providing confirmation of the CKM mechanism as the source of CP violation
in the Standard Model~\cite{Ab02, Au02}, among other notable discoveries. The B~factories also made prolific discoveries of conventional and 
exotic hadronic states, including the famous $X(3872)$, which is the most cited paper published by the Belle collaboration. Even ten years 
after data taking, Belle is regularly producing new results in hadron spectroscopy, with more than 350 papers published since the shutdown. With 
significant upgrades to the accelerator and detector at the KEK accelerator facility in Tsukuba, Japan, the Belle~II experiment inherits the rich 
physics program established at the B~factories to usher in the first Super~B~factory.

\begin{figure*}
\includegraphics[width=0.9\textwidth]{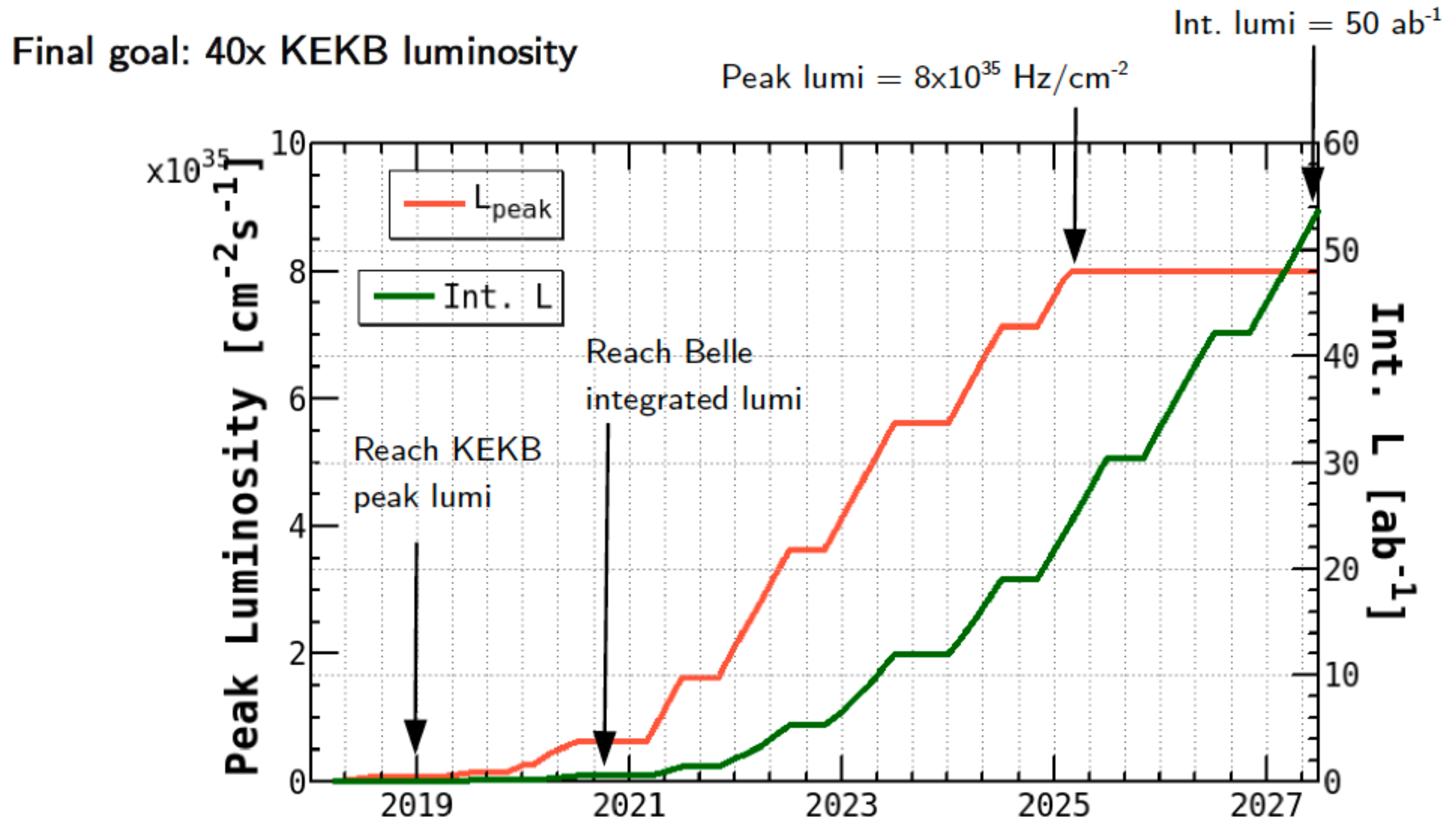}
\caption{The projected integrated (top) and instantaneous (bottom) luminosity for Belle II at SuperKEKB.}
\label{lumi}
\end{figure*}

The Belle~II physics program will leverage the unprecedented luminosity of the new SuperKEKB accelerator to collect a projected total data set of 
50~ab$^{-1}$ over the next decade, as shown in Figure~\ref{lumi}, enabling extremely precise measurements of suppressed flavor physics reactions 
that are sensitive to new physics. Belle II will also have access to a unique data set in which to study quarkonium-like states. To achieve the
unprecedented instantaneous luminosity expected from SuperKEKB, the accelerator uses the ``nano-beam'' scheme to squeeze the beam size
by an order of magnitude. Coupled with an increase in beam currents, the goal of SuperKEKB is to achieve an instantaneous luminosity about 
forty times that of KEKB.

To enable precise studies of CP violation and probe for new, interesting states and interactions, the Belle~II detector includes several significant
upgrades over the Belle detector. This includes a new silicon vertex detector, consisting of 2 layers of DEPFET pixels and 4 layers of 
double-sided silicon detectors, which will improve vertex resolution, even considering the reduced beam energy asymmetry. The helium-based 
central drift chamber has been expanded, which provides a longer lever-arm for charged particle tracking. To improve particle identification (PID)
performance, a sophisticated PID system was installed. This includes a time-of-propagation detector, made up of quartz bars, in the barrel region
and a proximity-focusing aerogel ring-imaging Cherenkov detector in the forward region. To cope with the much higher beam backgrounds at 
Belle~II, the CsI(Tl) electromagnetic calorimeter uses upgraded electronics with waveform sampling. The total expected data rate at the maximum
level-1 trigger rate of 30 kHz is over 30 GB per second. To cope with the enormous data sample, data processing, Monte Carlo production, and user
analysis are distributed over the world via the grid, following the LHC computing model. More details about the Belle~II detector are available in
Ref.~\cite{TDR}.

While the most high profile analyses expected from Belle~II focus on flavor physics, there are significant benefits for hadron spectroscopy at
B~factory experiments. In particular, Belle~II is a high resolution, hermetic detector with good PID capability and can efficiently reconstruct neutral
particles, which is challenging at LHCb. With a precise knowledge of the initial state, Belle~II has the capability to reconstruct a single resonance and 
explore the recoiling system, which is a significant source of particle discoveries, and can even use fully reconstructed events to measure absolute
branching fractions, which is essential for understanding XYZ states. Finally, Belle~II can access a wide variety of production mechanisms for 
hadronic states, from B~decays and initial-state radiation to two-photon and double charmonium production, in addition to quarkonium transitions.

Commissioning of the SuperKEKB accelerator was completed in two phases, the first in the spring of 2017 to study single beam backgrounds 
and the second during the spring and summer of 2018 to validate the ``nano-beam'' scheme responsible for the unprecedented instantaneous 
luminosity expected from SuperKEKB. The latter commissioning phase, called ``phase 2'', included first collisions of the SuperKEKB beams inside 
an incomplete version of the Belle II detector without most of the vertex detector (VXD), which is highly sensitive to beam backgrounds. A total of 
about 0.5 fb$^{-1}$ of data were collected during phase 2 and successful operation of the data acquisition system with all installed subsystems 
was achieved. The VXD was successfully installed in December of 2018, in preparation for data taking with the full Belle II detector during 
``phase 3'', which began in March of 2019. As of early June 2019, Belle~II has collected more than 5 fb$^{-1}$ and extensive detector performance
studies are underway in preparation for higher statistics data from later running.

\section{Prospects for XYZ at Belle~II}

With full Belle~II statistics, a copious number of interesting known states, such as the $X(3872)$, $Y(4260)$ and $Z(4430)$ among others, will be 
produced. This will allow precise determination of resonance parameters, such as a first determination of the absolute width of the $X(3872)$. It will 
also allow for searches for related states via transitions and amplitude analyses to determine the quantum numbers of the states.

One of the unique aspects of the XYZ program at Belle~II is the ability to study $Z$ states in both B~meson decays and in direct production via
initial-state radiation. This is particularly interesting because the $Z$ states discovered in B~meson decays appear to be much wider than their
counterparts in direct production. Higher statistics studies and searches for new production and decay mechanisms will be useful to clarify what
appears to be two distinct sets of $Z$~states.

Direct studies of $Y$ states are possible at electron-positron machines since their quantum numbers match those of the virtual photon produced in
the $e^{+}e^{-}$ collision. The lower energy symmetric collision data set collected by the BESIII experiment has yielded interesting results that 
suggest the $Y$ states such as the $Y(4260)$ have a much more complicated than anticipated lineshape~\cite{A17}, which may actually be 
composed of more than one state. The beam energies at Belle~II prevent such direct searches. However, the $Y$ spectrum is accessible via initial 
state radiation (ISR), in which one beam radiates a photon, bringing the collision energy into the charmonium region. With full statistics, the ISR data 
set collected by Belle~II will rival the direct production samples at BESIII. More importantly, while BESIII must change the beam energy and collect 
many distinct samples to study lineshapes, Belle~II will collect data at all $\sqrt{s}$, simultaneously. This will make lineshape studies much more 
accessible.

In addition to direct production and production in B~meson decays, a potential source of XYZ states is via hadronic transitions. With a precise 
knowledge of initial states, Belle~II can reconstruct a single particle and explore the recoil spectrum. This method has proven effective to discover
new states such as the $Z_{b}(10610)$ and $Z_{b}(10650)$~\cite{B12}. It is also possible for Belle~II to reconstruct a charmonium (or any -onium) 
state and explore the recoil spectrum. This double-charmonium production has also proven useful to study states such as the $X(3940)$~\cite{A072}. 
Similarly, two-photon production of charmonium states has produced interesting results on states like the $X(4350)$~\cite{S10}, which lacks 
experimental confirmation in the same spectrum produced in B~decays at LHCb~\cite{Aa17}.

Collectively, these various sources of XYZ state production provide Belle~II with a unique capability to study known quarkonium-like states and 
search for new ones. While most of the know states lie in the charmonium region, due to the relatively small samples collected above the 
$\Upsilon(4S)$, the beauty sector will undoubtedly prove a source of additional XYZ states. With the large samples above the $\Upsilon(4S)$ 
that Belle~II will collect, it will be important to compare and contrast the b- and c-quark systems to gain a deeper understanding of quarkonium-like
states.

While the meson spectrum receives most of the attention from experimentalists, due to its relative simplicity, Belle~II will also investigate exotic
quarkonium in the baryon spectrum. While it is more complicated, the baryon spectrum presents potential studies of exotic states even in the first
excited states. Notable examples include the $\Lambda(1405)$ and $N(1440)$. The excited spectrum is relatively poorly understood, with many 
missing states and multiple candidates for known states. The Belle experiment is still actively publishing baryon results and it will be a fruitful
area for convention and exotic quarkonium studies at Belle~II, which can measure quantum numbers for excited charmed baryons and use 
charmed baryon decays to search for excited baryons like the $\Xi(1620)$~\cite{S19}, which is itself an exotic candidate state.

\section{Summary}

Major upgrades at the KEK accelerator facility in Tsukuba, Japan represent the first super~B~factory experiment, hosting the Belle~II detector, 
which itself contains many upgrades and improvements over the Belle experiment. In addition to upgraded detector components and electronics,
the Belle~II software and analysis tools are superior to previous experiments. All of this work is directed toward enabling a rich physics program.
With the unprecedented luminosity expected from the SuperKEKB accelerator, the Belle~II experiment will make a significant impact on exotic
quarkonium spectroscopy.

\bigskip 

\end{document}